\documentclass[12pt]{iopart}

\usepackage{graphicx}
\usepackage{iopams}

\newcommand{\mb}{\mathbf}

\begin{document}

\title{Quantum analysis of the direct measurement of light waves}

\author{Pablo L Saldanha$^{1,2}$} 
\address{$^1$ Departamento de F\'isica, Universidade Federal de Pernambuco, 50670-901, Recife, PE, Brazil}
\address{$^2$ Departamento de F\'isica, Universidade Federal de Minas Gerais, Caixa Postal 702, 30161-970, Belo Horizonte, MG, Brazil}

\ead{saldanha@fisica.ufmg.br}

\begin{abstract}
In a beautiful experiment performed about a decade ago, Goulielmakis \textit{et al.} made a direct measurement of the electric field of light waves [Goulielmakis E \textit{et al.} 2004 {\em Science} \textbf{305} 1267--1269]. However, they used a laser source to produce the light field, whose quantum state has a null expectation value for the electric field operator, so how was it possible to measure this electric field? Here we present a quantum treatment for the $f$:$2f$ interferometer used to calibrate the carrier-envelope phase of the light pulses in the experiment. We show how the special nonlinear features of the $f$:$2f$ interferometer can change the quantum state of the electromagnetic field inside the laser cavity to a state with a definite oscillating electric field, explaining how the ``classical'' electromagnetic field emerges in the experiment.  We discuss that this experiment was, to our knowledge, the first demonstration of an absolute coherent superposition of different photon number states in the optical regime.
\end{abstract}



\maketitle

\section{Introduction}

In the experiment of Goulielmakis  \textit{et al.}
\cite{goulielmakis04} depicted in figure 1 they produce a  few-cycle
laser pulse with polarization in the $x$ direction that
co-propagates with an extreme ultraviolet (XUV) pulse. The XUV pulse
is generated by the same laser field that is going to be measured,
having a well-defined $z$ position in relation to the phase of the
laser field \cite{baltuska03}, and this relative position can be
further controlled. The pulses are focused in an atomic gas target
and the XUV pulse knocks electrons free by photoionization. Some of
these electrons propagate to an electron detector, suffering an
oscillating electric force due to the laser field from the moment of
their release. The kinetic energy of the electrons is measured by
the detector as a function of the $z$ position of the XUV pulse in
relation to the laser pulse. The variation of the electrons kinetic
energy depends on the integral of the electric field from the
instant of the photoionizations until the end of the laser pulse,
such that the laser electric field can be extracted from the data
\cite{goulielmakis04}.

Goulielmakis  \textit{et al.} used a laser source to produce the
light pulses in the experiment, and the quantum state of the
electromagnetic field of a single-mode laser can be written as
\cite{scully, mandel, molmer97, sanders03}
\begin{equation}\label{laser}
    \rho=\int_0^{2\pi} \frac{\rmd\phi}{2\pi} |\alpha\mathrm{e}^{\rmi\phi}\rangle_\mathrm{c\,c}\langle\alpha\mathrm{e}^{\rmi\phi}|=\mathrm{e}^{-|\alpha|^2}\sum_n\frac{|\alpha|^{2n}}{n!}|n\rangle_\mathrm{c\,c}\langle n|,
\end{equation}
where $|\alpha\rangle_\mathrm{c}$ represents a coherent state and $|n\rangle_\mathrm{c}$ a
number state for the electromagnetic field in the laser cavity mode. The
state can be written as a mixture of coherent states with amplitude
$|\alpha|$ and arbitrary phases or as a mixture of number states
with a Poissonian distribution. As it was discussed by M\o lmer
\cite{molmer97}, considering that the laser field is produced by
quantum transitions in an incoherently pumped gain medium, the
system dynamics forbids the field reduced density matrix to have a
coherent superposition of number states, which is a necessary
condition for having a nonzero expectation value for the electric
field operator \cite{scully,mandel}. So the quantum light field produced in a laser
cavity has a fundamentally undetermined electric field and cannot be
in a coherent state, as is usually assumed. To see why this is true, imagine that the (two-level) atoms of the cavity and the laser field resonant with the atomic transition form a closed system. If the system initially has $N$ excitations distributed among atoms and photons, it keeps the excitation number during the evolution, since that for each photon that is absorbed one atom makes the transition to the excited state and for each created photon one atom makes the transition to the fundamental sate. So a general quantum state after the evolution of this closed system is of the form $|\Psi\rangle_\mathrm{T}=\sum_n a_n|n\rangle_\mathrm{c}|\psi_{N-n}\rangle_\mathrm{at}$, where $|n\rangle_\mathrm{c}$ represents a cavity field state with $n$ photons and $|\psi_{N-n}\rangle_\mathrm{at}$ a state for the atoms with $N-n$ excitations distributed among them. If we compute the reduced state for the cavity field by tracing out the atomic degrees of freedom, we have $\rho=\sum_n |a_n|^2 |n\rangle_\mathrm{c\,c}\langle n|$, without coherences between different photon number states. To describe the complete dynamics of the system we must include an incoherent pumping and losses in the model, but these included terms obviously also cannot generate coherence between different photon number states. So in principle it is not possible that the reduced density matrix of the light state of a laser cavity has a nonzero expectation value for the electric field operator. However, as we will show in this work, the action of a $f:2f$ interferometer can generate an absolute coherence in the laser cavity field state and change this situation.

M\o lmer's work was inspired in the works that showed that the spatial superposition of two independent Bose-Einstein condensates can present  interference fringes even if each condensate is initially in a number state \cite{javanainen96,cirac96,wong96}. The spontaneous symmetry breaking in the system appears to the extent that the particles are detected in different regions of space. In an analogous way, M\o lmer showed that two independent light sources initially in number states combined in a beam splitter can present temporal interference \cite{molmer97}. The spontaneous symmetry breaking in the optical system appears to the extent that photons are detected in the different output ports of the beam splitter. Similar effects may also occur in other systems, such as in the interference of Cooper pairs emitted from independent superconductors \cite{iazzi10}. Recent works have
studied the influence of the fact that the laser field is not in a
coherent state in the interpretation of optics experiments
\cite{molmer97,sanders03,cable05}, in the realization of quantum
information protocols \cite{rudolph01,enk02,bartlett06}, in the
emergence of a photon-number superselection rule
\cite{sanders03,bartlett07}, in the meaning of phase coherence in
optical fields \cite{enk02,pegg09}, in optical quantum-state
tomography \cite{kawakubo10} and in quantum metrology
\cite{jarzyna12}.

\begin{figure}\begin{center}
  \includegraphics[width=10cm]{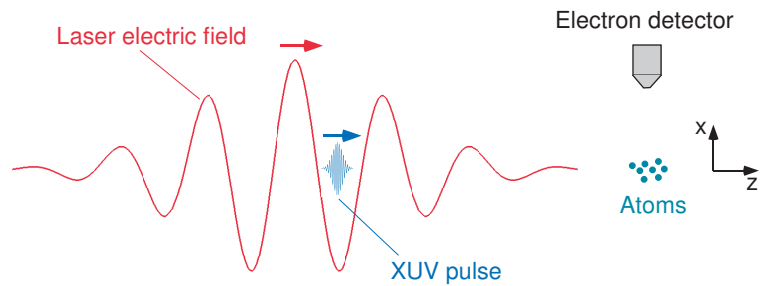}\\
  \caption{Schematic of the principle of the direct measurement of light waves  \cite{goulielmakis04}. A  few-cycle laser pulse with polarization in the $x$ direction co-propagates with an extreme ultraviolet (XUV) pulse. The pulses are focused in an atomic gas target and the XUV pulse knocks electrons free by photoionization. The kinetic energy of the electrons is measured by the electron detector as a function of the $z$ position of the XUV pulse in relation to the laser pulse, such that the laser electric field can be extracted from the data
\cite{goulielmakis04}.}
\label{fig1}
 \end{center}\end{figure}

Our primary motivation in this work is to describe how it was
possible for Goulielmakis  \textit{et al.} to produce  laser pulses
with a definite electric field like the one depicted in figure 1
starting with the state  (\ref{laser}) in the laser cavity, which has a null expectation value for the electric field operator. By
constructing a quantum treatment for the $f$:$2f$ interferometer \cite{jones00,apolonski00},
which was used to calibrate the carrier-envelope phase of the light
pulses in the experiment, we provide such a description. We show
that the nonlinear properties of the $f$:$2f$ interferometer are
capable of producing an absolute coherence in the laser cavity field
state, generating a quantum superposition of different photon number states, such that the expectation value of the electric field operator behaves like a classical wave. The present treatment gives a deeper understanding of the
process of phase control of few-cycle wave pulses using a $f$:$2f$ interferometer
\cite{jones00,apolonski00}, a very important tool that has been used
in a large variety of applications such as metrology
\cite{jones00,udem02,ludlow08}, spectroscopy
\cite{marian04,diddams07}, measurement and control of electronic
motion in the attosecond scale
\cite{goulielmakis07,kruger11,shafir12}, etc. \cite{krausz09}.

\section{Quantum treatment for the $f$:$2f$ interferometer}

In a single-mode pulsed regime, each of the photons in a laser cavity is in the same state
\begin{equation}\label{mode}
    \sum_j \gamma_j|f_j\rangle,\;\;\mathrm{with}\;\;f_j=j f_{\mathrm{rep}}+\delta\;\;\mathrm{and}\;\;\sum_j|\gamma_j|^2=1,
\end{equation}
in frequency mode,  $j$ assuming integer values, $f_{\mathrm{rep}}$ being the laser repetition rate and $\delta$ a frequency offset
\cite{jones00}. We will not consider the spatial or polarization
properties of the mode here. In this paper, kets with subscripts ``c'' like in (\ref{laser}) refer to the quantum state of the electromagnetic field in the laser cavity mode, while kets without subscripts like in (\ref{mode}) refer to monochromatic photon states. As it is expressed above, the laser cavity mode can be written as a linear superposition of monochromatic modes. A state with $m$ photons in this laser cavity mode
can be written as $|m\rangle_\mathrm{c}=(m!)^{-1/2}(\hat{b}_{\mathrm{c}}^\dag)^m
|\mathrm{vac}\rangle_\mathrm{c}$, where $|\mathrm{vac}\rangle_\mathrm{c}$ represents the
vacuum state for the electromagnetic field in the cavity and
$\hat{b}_{\mathrm{c}}^\dag\equiv\sum_j \gamma_j\hat{a}^\dag(f_j)$ the creation
operator for the laser mode, where $\hat{a}^\dag({f_j})$ is the
creation operator for the monochromatic mode with frequency $f_j$
\cite{titulaer66}.

When $\delta=0$ in  (\ref{mode}), all pulses exit the cavity with
the same (but undetermined) phase between the field oscillations and
the amplitude envelope, the so called carrier-envelope phase, while
for $\delta\neq0$ the carrier-envelope phase changes by
$2\pi\delta/f_{\mathrm{rep}}$ in each pulse \cite{jones00}. For this reason,
it was necessary to Goulielmakis \textit{et al.} to calibrate
$\delta$ to a fixed value in their experiment \cite{goulielmakis04}
using a $f$:$2f$ interferometer \cite{jones00, apolonski00} like the
simplified version depicted in figure 2(a). This was necessary because they had to collect data measuring the electric field of identical pulses in the experiment.  In the scheme of figure 2(a), each pulse leaves the
cavity through the partially reflective mirror M2, passing through
the photonic crystal fiber PCF that broadens the spectrum of the
field to more than one octave, as represented in figure 2(b). The
dichroic mirror DM reflects the frequency components above a
frequency $f'$ and transmits the frequencies below $f'$. The
transmitted field passes through a nonlinear BBO crystal aligned to
maximize the second harmonic generation from frequency $f_{x}$ to
$2f_{x}\approx f_{2x}$ as indicated in figure 2(b). The 50\% beam
splitter BS then combines the fields of both arms, that can have a
variable phase $\phi$ between the optical paths, and detectors D$_1$
and D$_2$ detect photons with frequencies around $f_{2x}$ selected
by the optical filters F$(f_{2x})$. To simplify the calculations, we
will consider that the neutral filter F$_0$ balances the
interferometer.

\begin{figure}\begin{center}
  \includegraphics[width=10cm]{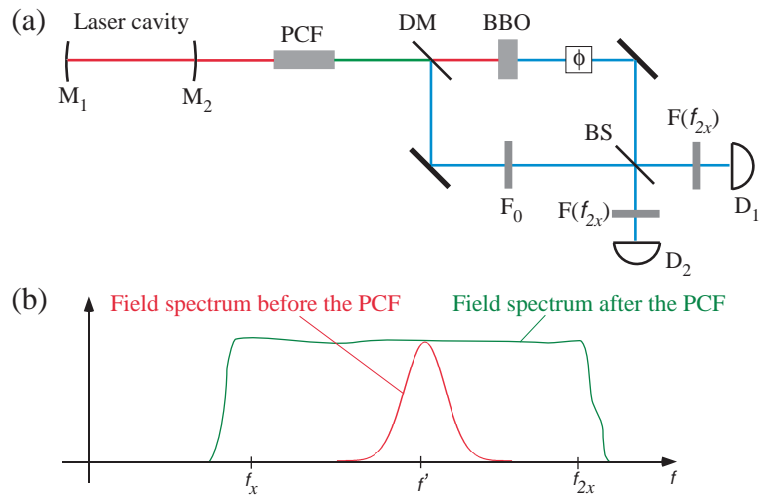}\\
  \caption{Simplified scheme of a $f$:$2f$ interferometer. (a) Interferometer setup. M$_1$, fully reflective mirror; M$_2$, partially reflective mirror; PCF, photonic crystal fiber that broadens the field spectrum; DM, dichroic mirror that reflects the higher frequency components and transmits the lower ones; BBO, nonlinear crystal for second harmonic generation; $\phi$, phase between the interferometer arms; BS, 50\% beam splitter; F($f_{2x}$), interference filter that transmits frequencies around $f_{2x}$, F$_0$, neutral filter; D$_1$ and D$_2$, detectors.
  (b) Illustration of the field spectrum broadening caused by the photonic crystal fiber (PCF).}
\label{fig2}
 \end{center}\end{figure}

Let us now see how the detection of one photon by detector D$_1$ in
figure 2(a) changes the quantum state of the cavity field. For simplicity, let us first consider
that the cavity field is initially in a pure number state  $|m\rangle_\mathrm{c}$ with $m$ photons in the
mode (\ref{mode}). By considering this state for the cavity, which is very distinct from the more realistic state (\ref{laser}), we will be able to discuss the influence of the photon measurements at the exits of the $f$:$2f$ interferometer on the cavity field state in a simpler manner keeping the principal features of the process, since the  state $|m\rangle_\mathrm{c}$ also has a null expectation value
for the electric field operator \cite{scully,mandel}. Latter we will discuss what changes in the process when we consider the initial state  (\ref{laser}) for the cavity field state. The detection
of one photon with frequency around $f_{2x}$ by D$_1$ at time $t$ is
associated to the action of the annihilation operator $\hat{A}_1(t)$
in the optical mode just before the detector. We can associate the
annihilation operator just before the detector to the annihilation
operator just after the PCF as $\hat{A}_1(t)=
\xi_1'\hat{a}_0(f_{2x})\rme^{-\rmi2\pi
f_{2x}t}+\xi_2'\hat{a}_0(f_{x})\hat{a}_0(f_{x})\rme^{\rmi(\phi-4\pi
f_xt)}$ with $\xi_1'$ and $\xi_2'$ being real constants. This occurs
because the second harmonic generation can be seen as the coherent
scattering of 2 photons into one, similarly to the case of
parametric downconversion \cite{saldanha11,saldanha11b} (that does
the opposite conversion). The detection could be the result of one
photon of frequency $f_{2x}$ that is reflected by the DM and
transmitted by the BS or of two photons of frequency $f_{x}$ that
are transmitted by the DM, converted into one photon of frequency
$2f_{x}$ by the BBO and reflected by the BS. The detector has a
large bandwidth, such that it cannot distinguish the frequencies
$2f_x$ and $f_{2x}$. Since these two situations are
indistinguishable, they must be coherently superposed, and the
corresponding probability amplitudes are proportional to $\xi_1'$
and $\xi_2'$ respectively.

The broadening of the field spectrum by the PCF indicated in Fig.
2(b) is related to the processes of four wave mixing and soliton
fission in the fiber \cite{ranka00, dudley06}.
Each photon at the entrance of the PCF can be scattered to any
frequency component at the exit of the PCF through energy exchange
with the other photons of the pulse in a coherent way. So we can
associate  $\hat{a}_0(f_{x})=\xi_1''\hat{b}_{\mathrm{c}}$ and
$\hat{a}_0(f_{2x})=\xi_2''\hat{b}_{\mathrm{c}}$, where $\hat{b}_{\mathrm{c}}$ is the
annihilation operator of the laser mode and we assume $\xi_1''$ and
$\xi_2''$ real (if they are not real, we can redefine the phase
$\phi$ in the interferometer of figure 2(a)). So, with the detection
of one photon by D$_1$ at time $t$ we have to apply an operator
proportional to $\xi_1\hat{b}_{\mathrm{c}}+\xi_2\rme^{\rmi(\phi-2\pi\delta
t)}\hat{b}_{\mathrm{c}}\hat{b}_{\mathrm{c}}$ in the cavity quantum field, with
$\xi_1\equiv \xi_1'\xi_1''$ and $\xi_2\equiv\xi_2'\xi_2''^2$, since
according to  (\ref{mode}) we have $2f_x-f_{2x}=\delta$. In an
analogous way, the detection of one photon by D$_2$ at time $t$ is
related to the application of the operator
$\xi_1\hat{b}_{\mathrm{c}}-\xi_2\rme^{\rmi(\phi-2\pi\delta t)}\hat{b}_{\mathrm{c}}\hat{b}_{\mathrm{c}}$
in the cavity quantum field. Thus if in the first pulse that exits
the cavity $n_1$ photons are detected by D$_1$ and $n_2$ photons are
detected by D$_2$ at time $t=0$ when the initial cavity state is a number state $|m\rangle_\mathrm{c}$, the cavity quantum field changes to
\begin{equation}\label{psi1}    |\Psi(m,n_1,n_2)\rangle_\mathrm{c}\propto\left[\xi_1\hat{b}_{\mathrm{c}}+\xi_2\rme^{\rmi\phi}\hat{b}_{\mathrm{c}}^2\right]^{n_1}\left[\xi_1\hat{b}_{\mathrm{c}}-\xi_2\rme^{\rmi\phi}\hat{b}_{\mathrm{c}}^2\right]^{n_2}|m\rangle_\mathrm{c}.
\end{equation}

In the Appendix it is shown that for  $n_1$ and $n_2$ large,
$m\gg n_1+n_2$  and a balanced interferometer, implying
$\xi_1\approx\sqrt{m}\xi_2$, the state  (\ref{psi1}) can be
written as
\begin{equation}\label{psi1a}
    |\Psi(m,n_1,n_2)\rangle_\mathrm{c}\propto\frac{1}{\sqrt{2}}\Big[|\Gamma(m,n,\phi+\pi n_2/n)\rangle_\mathrm{c}+|\Gamma(m,n,\phi-\pi n_2/n)\rangle_\mathrm{c}\Big],
\end{equation}
with $n\equiv n_1+n_2$ and
\begin{equation}\label{gamma}
    |\Gamma(m,n,\Phi)\rangle_\mathrm{c}\equiv\sum_k B({n,k}) \rme^{\rmi k\Phi}|m-n-k\rangle_\mathrm{c}
\end{equation}
with $B({n,k})\propto\exp\left[{-2(k-n/2)^2}/{n}\right]$.

The electric field operator at the position $\mb{r}=0$ inside the
laser cavity can be written in terms of an orthogonal set of modes
that includes the laser mode  (\ref{mode}): $\hat{E}(t)=\sum_l
v_l(t)\hat{b}_l+v_l^*(t)\hat{b}_l^\dag$ \cite{titulaer66}. For the
mode of  (\ref{mode}), we have
$v_c(t)=\sum_j\sqrt{hf_j/(2\varepsilon_0V)}\gamma_j\rme^{-i2\pi f_j
t}$, where $V$ is the effective mode volume, $\varepsilon_0$ the
permittivity of free space and $h$ the Planck constant. The states
$|\Gamma(m,n,\Phi)\rangle_\mathrm{c}$ from  (\ref{gamma}) have a classical
behavior, in the sense that they have a macroscopic expectation
value for the electric field operator. It is readily shown that for
the pulsed laser mode of  (\ref{mode}) with $\gamma_j$ real  we
have
\begin{eqnarray}\label{e}\nonumber
    &&_\mathrm{c}\langle\Gamma(m,n,\Phi)|\hat{E}(t) |\Gamma(m,n,\Phi)\rangle_\mathrm{c}=\sum_kB({n,k-1})\times\\
    &&\times B({n,k})\sum_j\sqrt{\frac{2hf_j(m-n-k)}{\varepsilon_0V}}\gamma_j\cos(2\pi f_j t+\Phi).
\end{eqnarray}
For large $n$, $B({n,k-1})\approx B({n,k})$ and the above
expectation value for the electric field has almost the same value
as the one for a coherent  state $|\alpha\rangle_\mathrm{c}$ with
$\alpha=\sqrt{m-3n/2}\,\mathrm{e}^{\rmi\Phi}$. According to (\ref{e}), $\Phi$ defines the phase between the underlying carrier
wave and the envelope in each pulse.

However, the state after the photon measurements in D$_1$ and D$_2$
is not yet a ``classical'' state of the form of  (\ref{gamma}),
but the quantum superposition of  (\ref{psi1a}). The reduction to
a ``classical'' state occurs with the detections of photons from the
second pulse  at time $\tau=1/f_{\mathrm{rep}}$, $\tau$ being the interval
between two consecutive pulses. For the states of  (\ref{gamma})
with large $n$, $m\gg n$ and  $\xi_1\approx\sqrt{m}\xi_2$, the
probability of photon detection  by D$_1$ or D$_2$ at time
$\tau=1/f_{\mathrm{rep}}$ is proportional to \cite{mandel}
\begin{eqnarray}\label{probd1}\nonumber
    &&_\mathrm{c}\langle\Gamma| [\xi_1\hat{b}_{\mathrm{c}}^\dag\pm\xi_2\hat{b}_{\mathrm{c}}^{\dag2}\rme^{-\rmi(\phi-2\pi\delta\tau)}][\xi_1\hat{b}_{\mathrm{c}}\pm\xi_2\hat{b}_{\mathrm{c}}^2\rme^{\rmi(\phi-2\pi\delta\tau)}] |\Gamma\rangle_\mathrm{c}\propto \\ && \propto 1\pm\cos(\Phi+\phi-2\pi\delta/f_{\mathrm{rep}}),
\end{eqnarray}
with the plus (minus) sign referring to D$_1$ (D$_2$). So each state
$|\Gamma(m,n,\Phi)\rangle_\mathrm{c}$ in the superposition of  (\ref{psi1a})
predict different values for the  numbers of photons detected by
D$_1$ and D$_2$  in the second pulse. So we have a 50\% chance of
having the values predicted by the state $|\Gamma(m,n,\phi+\pi
n_2/n)\rangle_\mathrm{c}$ and the field is reduced to a state closely related
to this one and a  50\% chance of having the values predicted by the
state $|\Gamma(m,n,\phi-\pi n_2/n)\rangle_\mathrm{c}$ and a similar reduction
of the state. In any case the final quantum state is a ``classical''
state of the form of  (\ref{gamma}) and the subsequent laser
pulses that exit the cavity will have a definite oscillating
electric field.

We would like to emphasize a crucial difference between our treatment  and previous treatments that describe the emergence of a
relative coherence between two independent light sources through the
interference  and  measurement of the resultant fields
\cite{molmer97,sanders03,cable05}. In these previous works, the
relative coherence is a consequence of the fact that there is a
fundamental indistinguishability about from which source each of the
detected photons come, and the superposition of the two situations
generates an entangled state between the two sources. The
expectation value of the electric field in these cases is zero for
both sources at any time if it was zero at the beginning
\cite{molmer97,sanders03,cable05}. This is the same logic as in the interference of two independent Bose-Einstein condensates \cite{javanainen96,cirac96,wong96}. In the present case, what happens
is that there is a fundamental indistinguishability for each
detected photon about if it is the consequence of the conversion of
one or of two photons from the laser cavity field in the $f:2f$
interferometer of figure 2(a). The coherent superposition of these two
situations creates a coherent superposition of different number
states in the laser cavity field, generating a nonzero expectation
value for the electric field operator even if it was zero at the
beginning. We conclude that the special nonlinear properties of the $f:2f$
interferometer are capable of generating an absolute coherence in
the system.

According to  (\ref{probd1}), the intensity measured by D$_1$ in the N-th pulse obeys
\begin{equation}\label{intensity}
    I_1(N)\propto1+\cos[\Phi+\phi_N-(N-1)2\pi\delta/f_\mathrm{rep}],
\end{equation}
$\phi_N$ being the phase between the interferometer arms during the
N-th pulse. This interference pattern with controllable $\phi_N$ can
thus be used to determine $\delta$ from  (\ref{mode}). Our
treatment corresponds to a quantum description of the
carrier-envelope phase control of optical pulses using a $f$:$2f$ interferometer classically
described in  \cite{jones00, apolonski00}. As the laser is a
quantum source of radiation, since it cannot be described using
classical physics, a quantum treatment for the phenomenon is more
accurate, as it evidences the important role of the external
measurements on the construction of the laser cavity field quantum
state, changing it from a state with a fundamentally undetermined
electric field to a state with a definite oscillating electric
field.

\section{Analysis of the Goulielmakis \textit{et al.} experiment \cite{goulielmakis04}}

To perform the measurements of the electric field of light pulses in
\cite{goulielmakis04},  Goulielmakis \textit{et al.} fixed the frequency offset $\delta$ from  (\ref{mode})
using a $f$:$2f$ interferometer such that after every $M$ pulses the
carrier-envelope phase was the same as in the first pulse, and
selected identical pulses  to be measured. In this process, the
carrier-envelope phase acquires a definite but random value, as we have discussed, that
was determined by measuring the spectrum of XUV pulses emitted by
atoms subjected to the laser field \cite{baltuska03}. Then, by
adjusting some parameters of the laser cavity, they could adjust the
carrier-envelope phase to the desired value and proceed to measure
the electric field of identical light pulses. It is only after this calibration procedure that they perform the measurements described in figure \ref{fig1}  \cite{goulielmakis04}. As we have discussed, during the calibration process the laser field quantum state evolves from a state with a fundamentally undetermined electric field to a state with a definite oscillating electric field, that could then be measured following the procedure described in figure \ref{fig1}.

It is important to stress that the processes of determining the
carrier-envelope phase of the pulses, generating the attosecond XUV
probes and measuring the kinetic energy of the probe electrons in
general also change the laser cavity field quantum state and may
increase its degree of coherence.
But since the objective of the present work is to describe how a
nonzero expectation value for the electric field emerges in the
experiment \cite{goulielmakis04} in the first place, which is caused
by the $f$:$2f$ interferometer as described before, we will not
present a quantum treatment for these other processes here. The fact
that the $f:2f$ interferometer generates an absolute coherence in
the laser field can be confirmed simply by the fact that
interference is observed. Note that if we have a number state or an
incoherent combination of number states in  (\ref{probd1}), the
prediction for the intensity measured by D$_1$ in Eq.
(\ref{intensity}) would not depend on time or on the phase $\phi$ of
the interferometer. Since the situations that are superposed at the
exit of the interferometer are related to the annihilation of one
and of two photons from the field at the entrance of the PCF, there
is interference only if this field  has a quantum superposition of
different number states. For this reason, in the first pulse that
exits the cavity it is not possible to predict the intensities
measured by D$_1$ and D$_2$. However, as we have shown, after the
measurements of a couple of pulses coherence is generated in the
cavity field and the interference curve can be observed.

\section{Some considerations}

Up to now we have considered an initial pure state $|m\rangle_\mathrm{c}$ for
the cavity field and disregarded optical losses in the processes and
the optical gain in the cavity. But we can see that for a different
initial  number state the expectation value of the electric field in
 (\ref{e}) only changes its amplitude. So with an initial state
of the form of  (\ref{laser}) (that keeps this form with the
inclusion of losses and gain) there will be only a larger
uncertainty on the amplitude of the oscillating wave. The phase
$\Phi$ and the classical properties of the state will be essentially
the same. This behavior is analogous to the localization of the
relative phase between two optical modes through the interference
and measurement of photons from these modes, that is roughly the
same for pure and mixed states \cite{cable05}.

Many factors can decrease the visibility of the interference curve
of  (\ref{intensity}), such as incoherent processes in the PCF,
an unbalanced interferometer and misalignments. In this case, the
photon counts in D$_1$ and D$_2$ will be always equal to or higher
than a minimum value $n_\mathrm{min}$ in each pulse. The difference
of the treatment in this case is that we must replace $n_1$ and
$n_2$ in  (\ref{psi1}) by $n_1-n_\mathrm{min}$ and
$n_2-n_\mathrm{min}$. This occurs because in this new situation
these $2n_\mathrm{min}$ photons that are detected by D$_1$ and D$_2$
do not contribute to the construction of a quantum superposition of
different number states for the quantum field, so they can be
considered as optical losses.

\section{Conclusion}

By presenting a quantum treatment for the $f$:$2f$ interferometer,
we have shown how its special nonlinear properties can change the quantum state of a laser cavity
field from a state with null expectation value for the electric
field to a state that has a ``classical'' macroscopically  oscillating electric
field, describing the role of quantum measurements on the process. The emergence of this ``classical'' field with an absolute coherence in the laser source is essential for the measurement of the electric field of light in the Goulielmakis \textit{et al.} experiment \cite{goulielmakis04}.

We would like to finish by saying that M\o lmer's conjecture in his
seminal 1997 paper \cite{molmer97} that ``... optical coherences,
i.e., quantum-mechanical coherences between states separated by Bohr
frequencies in the optical regime, do not exist in optics
experiments'' is not valid anymore. As we discussed here, the use of a $f:2f$ interferometer can generate an absolute coherence in an initially incoherent source. In this sense, the experiment of
Goulielmakis \textit{et al.} \cite{goulielmakis04} was not only the
first one to directly measure the electric field of light, but also
the first one to clearly demonstrate the construction of a light
state with a definite oscillating electric field, characterizing an
absolute coherent superposition of different photon number states,
without an optical phase reference \cite{bartlett06,bartlett07}.

\ack

The author acknowledges Nicim Zagury, Jos\'e R. Rios Leite, Edilson
L. Falc\~ao-Filho, Marcelo Fran\c ca Santos and Marcelo Terra Cunha
for useful discussions. This work was supported by the Brazilian
agencies CNPq and FACEPE.

\appendix
\setcounter{section}{1}
\section*{Appendix}

Here we show how the state  (\ref{psi1}) can be
approximated to the state  (\ref{psi1a}) for $n_1$ and $n_2$
large, $m\gg n_1+n_2$ and $\xi_1\approx\sqrt{m}\xi_2$. Making a
polynomial expansion in  (\ref{psi1}) and using the fact that
$\hat{b}_{\mathrm{c}}|m\rangle_\mathrm{c}=\sqrt{m}|m-1\rangle_\mathrm{c}$, we can write
\begin{eqnarray}\nonumber
    |\Psi\rangle_\mathrm{c}&\propto&\sum_{p=0}^{n_1}\sum_{q=0}^{n_2}(^{n_1}_{p})(^{n_2}_{q})\sqrt{m(m-1)...(m-n-k+1)}\times\\
    &&\times\xi_1^{n-k}\xi_2^k\,\rme^{\rmi k\phi}(-1)^{q}|m-n-k\rangle_\mathrm{c}
\end{eqnarray}
with $(^{a}_{b})\equiv a!/[b!(a-b)!]$, $k\equiv p+q$ and $n\equiv n_1+n_2$. For $m\gg n$ and very large, we have $\sqrt{m(m-1)...(m-n-k+1)}\approx m^{k/2}$ and for a balanced interferometer $[\xi_2\sqrt{m}/\xi_1]^k\approx1$. Under these approximations, we can write
\begin{equation}
    |\Psi\rangle_\mathrm{c}\propto\sum_{p=0}^{n_1}\sum_{q=0}^{n_2}(^{n_1}_{p})(^{n_2}_{q})\rme^{\rmi k\phi}(-1)^{q}|m-n-k\rangle_\mathrm{c}.
\end{equation}
For large $n_1$ and $n_2$, we can write $(^{n_1}_{p})\propto \exp[-2(p-n_1/2)^2/n_1]$, $(^{n_2}_{q})\propto \exp[-2(q-n_2/2)^2/n_2]$ and extend the summations in $p$ and $q$ from $-\infty$ to $\infty$. Using these approximations and changing the summation index from $p$ to $k$, after some algebra we obtain
\begin{eqnarray}\label{a2}\nonumber
      && |\Psi\rangle_\mathrm{c}\propto \sum_k\sum_q\exp\left[ \frac{-2n(q-kn_2/n)^2}{n_1n_2} \right]\rme^{\rmi\pi q} \times\\
      &&\times  \exp\left[\frac{-2(k-n/2)^2}{n}\right]\rme^{\rmi k\phi}|m-n-k\rangle_\mathrm{c},
\end{eqnarray}
where we used $(-1)^{q}=\rme^{\rmi\pi q}$.

For large $\sigma$, the complex Fourier series expansion of the function $f(x)=\exp(-\sigma^2x^2/2)\cos(\mu x)$, $f(x)=\sum_q c_q\mathrm{e}^{\rmi qx}$, has coefficients
\begin{eqnarray}\nonumber
    c_q&=&\frac{1}{4\pi}\int_{-\pi}^\pi \rme^{-\sigma^2x^2/2}\left[\rme^{\rmi(q-\mu)x}+\rme^{\rmi(q+\mu)x}\right]\rmd x \\ &\approx& \frac{1}{2\sqrt{2\pi}\sigma}\left[\rme^{-(q-\mu)^2/(2\sigma^2)}+\rme^{-(q+\mu)^2/(2\sigma^2)}\right],
\end{eqnarray}
since the integrals can be extended from $-\infty$ to $\infty$ for large $\sigma$. Rearranging the terms in the Fourier expansion, we can write $f(\pi)$ as
\begin{equation}
    \exp(-\sigma^2\pi^2/2)\cos(\mu \pi)\propto\sum_q \exp\left[\frac{-(q-\mu)^2}{2\sigma^2}\right]\rme^{\rmi\pi q}.
\end{equation}
Substituting the above expression in  (\ref{a2}) with $\mu= kn_2/n$ and $\sigma^2= n_1n_2/(4n)$ we obtain
\begin{equation}
|\Psi\rangle_\mathrm{c}\propto \sum_k \exp\left[\frac{-2(k-n/2)^2}{n}\right]
  \cos\left[\frac{k n_2 \pi}{n} \right]\rme^{\rmi k\phi}|m-n-k\rangle_\mathrm{c}.
\end{equation}
It can be readily shown that the above equation can be written as
 (\ref{psi1a}).



\end{document}